\RequirePackage{fix-cm}
\documentclass[smallextended]{svjour3}
\smartqed
\usepackage{amsfonts,amssymb,amsmath}
\usepackage{graphics,graphicx,epsfig}
\usepackage[dvipsnames]{xcolor}
\usepackage[normalem]{ulem}
\usepackage{comment}
\usepackage{epstopdf}
\usepackage{subfigure}
\usepackage{marvosym}
\usepackage{hyperref}
\usepackage{geometry}
\usepackage{booktabs}
\usepackage{cite}
\usepackage [latin1]{inputenc}

\def\be{\begin{eqnarray}}
\def\ee{\end{eqnarray}}
\def\ba{\begin{array}{l}}
\def\ea{\end{array}}

\begin{document}
\title{Detection of multipartite entanglement based on Heisenberg-Weyl representation of density matrices}
\author{Hui Zhao$^1$ \and Yu Yang $^1$  \and Naihuan Jing$^{2}$  \and Zhi-Xi Wang$^3$ \and Shao-Ming Fei$^{3,4}$}
\institute{ \Letter~~corresponding author:Hui Zhao \at
                     zhaohui@bjut.edu.cn \and
              \Letter~~Yu Yang \at
                     1621740794@qq.com \and
            \Letter~~    Naihuan Jing\at
                    ~~~ jing@math.ncsu.edu\and
              \Letter~~     Zhi-Xi Wang\at
                     ~~~wangzhx@cnu.edu.cn\and
              \Letter~~     Shao-Ming Fei\at
                     ~~~feishm@cnu.edu.cn
           \and
 \at 1 Department of Mathematics, Faculty of Science, Beijing University of Technology, Beijing 100124, China\\
 \at 2 Department of Mathematics, North Carolina State University, Raleigh, NC 27695, USA\\
 \at 3 School of Mathematical Sciences, Capital Normal University, Beijing 100048, China\\
 \at 4 Max-Planck-Institute for Mathematics in the Sciences, Leipzig 04103, Germany
}
%%%%%%%%%%%%%%%%%%%%%%%%%%%%%%%%

\maketitle
\begin{abstract}
\baselineskip18pt
We study entanglement and genuine entanglement of tripartite and four-partite quantum states by using Heisenberg-Weyl (HW) representation of density matrices. Based on the correlation tensors in HW representation, we present criteria to detect entanglement and genuine tripartite and four-partite entanglement. Detailed examples show that our method can detect more entangled states than previous criteria.

\keywords{Heisenberg-Weyl representation \and Genuine entanglement \and  Correlation tensor}
\end{abstract}

\section{Introduction}
One of the most distinguished features of quantum mechanics is the
quantum entanglement which has wide applications in various areas
%Quantum entanglement is useful in different fields,
of quantum information processing such as quantum cryptography \cite{1EA},
quantum secure communication \cite{2BW} and quantum channel protocols \cite{3BBJ}.
Many criteria have been given to detect quantum entanglement, such as positive partial transpose criterion \cite{4HHH}, entanglement witness \cite{5B} and realignment of density matrices \cite{6CW,7R,8RO,9LWSCF}. Genuine multipartite entanglement based on the lower bounds of concurrence was studied in \cite{10CMCS}. In \cite{11ZFFW} inequalities for detecting
quantum entanglement of bipartite quantum systems and necessary conditions of
separability for mixed states were given. A witness test to detect genuine multipartite
entanglement of bound entangled states was presented in \cite{12HS}.
New separability criteria for bipartite and multipartite quantum states based on the Bloch
representation of density matrices have been derived in \cite{13SYLF}.
Separability of bipartite quantum systems was discussed in terms of Bloch representation in \cite{14V}. In \cite{15BK}, separability for different classes of quantum states in multipartite systems was studied.

While the Bloch representation of density matrices uses three kinds of generators of special unitary Lie group $SU(d)$, the Heisenberg-Weyl (HW) representation has only one type of generators. The obvious conciseness of the HW representation suggests possibility to simplify some of the calculations related to density matrices.
In \cite{16AEHK}, generalized Pauli matrices based on the HW operators were introduced and a criterion to detect entanglement by the bounds of the sum of expectation values of any set of anti-commuting observables was given. Moreover, separability criteria in terms of the HW observable basis for multipartite quantum systems were presented in \cite{17CCZ}.

In this paper, we study the genuine multipartite entanglement (GME) of tripartite and four-partite quantum states by exploiting the HW representation of density matrices. This paper is organized as follows. In Section 2, we use the correlation tensor to construct a special matrix to give a criterion of genuine tripartite entanglement according to the HW representation. An example is given to show that our result can detect more entanglement. In Section 3, a new criterion of genuine four-partite entanglement is presented. By a detailed example, our results are seen to outperform some preciously available results. In Section 4, conclusions are given and results are summarized.

\section{Detection of Genuine Tripartite Entanglement} \label{The bound}
We first consider the GME for tripartite quantun states. Let $d$ be the
dimension of the underlying space. The HW operators are defined as follows \cite{15BK}:
\begin{equation}
\begin{split}\label{1}
\mathcal{D}(l,m)=\sum_{k=0}^{d-1}e^{\frac{2i\pi kl}{d}}|k\rangle\langle(k+m)\, mod\, d|,\quad l,m=0,1,...,d-1.
\end{split}
\end{equation}
The HW observable basis is given by
\begin{equation}
\begin{split}\label{2}
\mathcal{Q}(l,m)=\mathcal{X}\mathcal{D}(l,m)+\mathcal{X}^{\ast}\mathcal{D}^{\dagger}(l,m),\quad \mathcal{X}=\frac{(1\pm i)}{2},
\end{split}
\end{equation}
which satisfies the following relations,
\begin{equation}
\begin{split}\label{3}
\mbox Tr\{\mathcal{Q}(l,m)\mathcal{Q}(l',m')\}=d\delta_{l,l'}\delta_{m,m'}.
\end{split}
\end{equation}
When $d=2$ the HW observable basis $\mathcal{Q}(0,1)$, $\mathcal{Q}(1,0)$ and $\mathcal{Q}(1,1)$ give rise to the standard Pauli matrices.

Let $H_n^{d_n}$ ($n=1,2,3$) be $d_n$-dimensional Hilbert spaces and $I$ the $d_{n}\times d_{n}$ identity matrix. Denote $\|\cdot\|$ the Hilbert-Schmidt norm (Frobenius norm) and $\|\cdot\|_{tr}$ the trace norm defined by $\|A\|_{tr}=\sum_{i}\sigma_{i}=\mbox Tr\sqrt{A^{\dagger}A}$, $A\in\mathbb{R}^{m\times n}$, where $\sigma_{i}$ $(i=1,2,\cdot\cdot\cdot, \mbox {min}\{m,n\})$ are the singular values of the matrix $A$ arranged in descending order. A state $\rho\in H_1^{d_1}$ can be expressed as  $\rho=\frac{1}{d_{1}}I+\frac{1}{2}\sum t_{lm}\mathcal{Q}(l,m)$, where $t_{lm}=\frac{2}{d_{1}}\mbox Tr\{\rho \mathcal{Q}(l,m)\}$, $l,m=0,...,d_{1}-1,(l,m)\neq (0,0)$. Let $T^{(1)}$ be the column vector with entries $t_{lm}$. For any pure state, it is known that $\|T^{(1)}\|^{2}=\frac{4(d_1-1)}{d_1^{2}}$ \cite{17CCZ}.

A bipartite state $\rho \in \mbox {End}(H_{1}^{d_{1}}\otimes H_{2}^{d_{2}})$ can be expressed as
\begin{equation}
\begin{split}\label{5}
\rho=&\frac{1}{d_{1}d_{2}}I_{d_1}\otimes I_{d_2}+\sum_{f=1}^2\frac{d_{f}}{2d_{1}d_{2}}\sum_{(l_{f},m_{f})\neq (0,0)}t_{l_{f}m_{f}}^{f}\mathcal{Q}^{f}(l_{f},m_{f})\\
&+\frac{1}{4}\sum_{(l_{1},m_{1}),(l_{2},m_{2})\neq (0,0)}t_{l_{1}m_{1}l_{2}m_{2}}^{12}\mathcal{Q}^{1}(l_{1},m_{1})\mathcal{Q}^{2}(l_{2},m_{2}),
\end{split}
\end{equation}
where $\mathcal{Q}^1(l_{f},m_{f})=\mathcal{Q}(l_{f},m_{f})\otimes I$, $\mathcal{Q}^2(l_{f},m_{f})=I\otimes\mathcal{Q}(l_{f},m_{f})$, $t_{l_{f}m_{f}}^{f}=\frac{2}{d_{f}}\mbox Tr\{\rho \mathcal{Q}^{f}(l_{f},m_{f})\}$. We have $\|T^{(f)}\|^{2}=\sum\limits_{(l_{f},m_{f})\neq (0,0)}(t_{l_{f}m_{f}}^{f})^{2}$ and $\|T^{(12)}\|^{2}=\sum\limits_{(l_{1},m_{1}),(l_{2},m_{2}))\neq (0,0)}(t_{l_{1}m_{1}l_{2}m_{2}}^{12})^{2}$.

For any tripartite state $\rho \in \mbox {End}(H_{1}^{d_{1}}\otimes H_{2}^{d_{2}}\otimes H_{3}^{d_{3}})$, we have
\begin{equation}
\begin{split}\label{6}
\rho=&\frac{1}{d_{1}d_{2}d_{3}}I_{d_1}\otimes I_{d_2}\otimes I_{d_3}+\sum_{f=1}^{3}\frac{d_{f}}{2d_{1}d_{2}d_{3}}\sum_{(l_{f},m_{f})\neq (0,0)}t_{l_{f}m_{f}}^{f}\mathcal{Q}^{f}(l_{f},m_{f})\\
&+\sum_{1\leq f<g\leq 3}\frac{d_{f}d_{g}}{4d_{1}d_{2}d_{3}}\sum_{(l_{f},m_{f}),(l_{g},m_{g})\neq (0,0)}t_{l_{f}m_{f}l_{g}m_{g}}^{fg} \mathcal{Q}^{f}(l_{f},m_{f})\mathcal{Q}^{g}(l_{g},m_{g})\\
&+\frac{1}{8}\sum_{(l_{1},m_{1}),(l_{2},m_{2}),(l_{3},m_{3})\neq (0,0)}t_{l_{1}m_{1}l_{2}m_{2}l_{3}m_{3}}^{123}\mathcal{Q}^{1}(l_{1},m_{1}) \mathcal{Q}^{2}(l_{2},m_{2})\mathcal{Q}^{3}(l_{3},m_{3}),
\end{split}
\end{equation}
where $\mathcal{Q}^{f}(l_{f},m_{f})$
%($(f),(g)$ represents the position of $\mathcal{Q}(l_{f},m_{f})$, $\mathcal{Q}(l_{g},m_{g})$ in the tensor product respectively)
stands for that the operator $\mathcal{Q}(l_{f},m_{f})$ acts on the $f$th space and
identity acts on the remaining spaces, the coefficients $t_{l_{f}m_{f}}^{f}=\frac{2}{d_{f}}\mbox Tr\{\rho \mathcal{Q}^{f}(l_{f},m_{f})\}$,
$t_{l_{f}m_{f}l_{g}m_{g}}^{fg}=$ $\frac{4}{d_{f}d_{g}}\mbox Tr\{\rho \mathcal{Q}^{f}(l_{f},m_{f})\mathcal{Q}^{g}(l_{g},m_{g})\}$ and
$t_{l_{1}m_{1}l_{2}m_{2}l_{3}m_{3}}^{123}=\frac{8}{d_{1}d_{2}d_{3}}\mbox Tr\{\rho \mathcal{Q}^{1}(l_{1},m_{1})\mathcal{Q}^{2}(l_{2},m_{2}) \mathcal{Q}^{3}(l_{3},m_{3})\}$. Let $T^{(f)}$, $T^{(fg)}$ and $T^{(123)}$ be the column vectors with entries $t_{l_{f}m_{f}}^{f}$, $t_{l_{f}m_{f}l_{g}m_{g}}^{fg}$ and $t_{l_{1}m_{1}l_{2}m_{2}l_{3}m_{3}}^{123}$, respectively. We have
\begin{align*}
\|T^{(f)}\|^{2}&=\sum\limits_{(l_{f},m_{f})\neq (0,0)}(t_{l_{f}m_{f}}^{f})^{2},\\
\|T^{(fg)}\|^{2}&=\sum\limits_{(l_{f},m_{f}),(l_{g},m_{g})\neq (0,0)}(t_{l_{f}m_{f}l_{g}m_{g}}^{fg})^{2},\\
\|T^{(123)}\|^{2}&=\sum\limits_{(l_{1},m_{1}),(l_{2},m_{2}),(l_{3},m_{3})\neq (0,0)}(t_{l_{1}m_{1}l_{2}m_{2}l_{3}m_{3}}^{123})^{2}.
\end{align*}

\begin{lemma}\label{lemma:1} For any pure state $\rho=|\phi\rangle\langle\phi| \in H_{1}^{d_{1}}\otimes H_{2}^{d_{2}}$ we have
\begin{eqnarray}\label{7}
\|T^{(12)}\|^{2}\leq \frac{16(d_{2}^{2}-1)}{d_{1}d_{2}^{3}}.
\end{eqnarray}
\end{lemma}
{\it Proof}~~
Since $\rho$ is pure, we have $\frac {d_1}{4}\|T^{(1)}\|^2-\frac {d_2}{4}\|T^{(2)}\|^2=\frac1{d_2}-\frac1{d_1}$ by using $\mbox Tr\rho_{1}^{2}=\mbox Tr\rho_{2}^{2}$,
where $\rho_{1}$ and $\rho_{2}$ are the reduced density operators on $H_{1}^{d_{1}}$ and $H_{2}^{d_{2}}$, respectively. From (\ref{5}) we have
\begin{equation}
\begin{split}\label{8}
Tr\rho^{2}=&\frac{1}{d_{1}d_{2}}+\frac{d_{1}}{4d_{2}}\sum_{(l_{1},m_{1})\neq (0,0)}t_{l_{1}m_{1}}^{2}+\frac{d_{2}}{4d_{1}}\sum_{(l_{2},m_{2})\neq (0,0)}t_{l_{2}m_{2}}^{2}\\
&+\frac{d_{1}d_{2}}{16}\sum_{(l_{1},m_{1}),(l_{2},m_{2})\neq(0,0)}t_{l_{1}m_{1}l_{2}m_{2}}^{2}\\
=&1.
\end{split}
\end{equation}
Therefore,
\begin{equation}
\begin{split}\label{9}
\|T^{(12)}\|^{2}&=\sum_{(l_{1},m_{1}),(l_{2},m_{2})\neq(0,0)}t_{l_{1}m_{1}l_{2}m_{2}}^{2}\\
&=\frac{16(d_{2}^{2}-1)}{d_{1}d_{2}^{3}}-\frac{4(d_{1}+d_{2})}{d_{1}^{2}d_{2}}\sum_{(l_{2},m_{2})\neq (0,0)}t_{l_{2}m_{2}}^{2}\\
&\leq \frac{16(d_{2}^{2}-1)}{d_{1}d_{2}^{3}}.
\end{split}
\end{equation}
\qed

\begin{lemma}\label{lemma:2}
Let $\rho\in H_1^{d_1}\otimes H_2^{d_2}\otimes H_3^{d_3}$ be a pure state such that $d_{i}d_{j}\geq d_{k},ijk \in \{123,132,231\}$. Then
\begin{eqnarray}\label{10}
\|T^{(123)}\|^2\leq \frac{64[d_{1}d_{2}d_{3}(d_{1}d_{2}d_{3}+2)-(d_{1}^{2}d_{2}^{2}
+d_{1}^{2}d_{3}^{2}+d_{2}^{2}d_{3}^{2})]}{d_{1}^{3}d_{2}^{3}d_{3}^{3}}.
\end{eqnarray}
\end{lemma}
{\it Proof}~~
Since $\rho$ is a pure state, we have $\mbox Tr(\rho^2)=1$ and $\mbox Tr(\rho_{i_1}^2)=Tr(\rho_{i_2i_3}^2)$, where $\rho_{i_1}$ ($i_1=1, 2, 3$) and $\rho_{i_2i_3}$ ($1\leq i_2<i_3\leq3$) are the reduced density operators of $\rho$ with respect to the subsystems $H_{d_{i_1}}$ and $H_{d_{i_2}d_{i_3}}$, respectively.
Therefore,
\begin{equation}
\begin{split}\label{11}
\|T^{(123)}\|^2=&\frac{64(d_{1}d_{2}d_{3}-1)}{d_{1}^{2}d_{2}^{2}d_{3}^{2}}-16(\frac{1}{d_{2}^{2}d_{3}^{2}}\|T^{(1)}\|^{2}+\frac{1}{d_{1}^{2}d_{3}^{2}}\|T^{(2)}\|^{2}
+\frac{1}{d_{1}^{2}d_{2}^{2}}\|T^{(3)}\|^{2})\\
&-4(\frac{1}{d_{3}^{2}}\|T^{(12)}\|^2+\frac{1}{d_{2}^{2}}\|T^{(13)}\|^2+\frac{1}{d_{1}^{2}}\|T^{(23)}\|^2)\\
\leq&\frac{64[d_{1}d_{2}d_{3}(d_{1}d_{2}d_{3}+2)-(d_{1}^{2}d_{2}^{2}+d_{1}^{2}d_{3}^{2}+d_{2}^{2}d_{3}^{2})]}{d_{1}^{3}d_{2}^{3}d_{3}^{3}}\\
&-\frac{16}{d_{1}^{2}d_{2}^{2}d_{3}^{2}}[(d_{2}d_{3}-d_{1})\cdot \|T^{(1)}\|^{2}+(d_{1}d_{3}-d_{2})\cdot \|T^{(2)}\|^{2}\\
&+(d_{1}d_{2}-d_{3})\cdot\|T^{(3)}\|^{2}]\\
\leq &\frac{64[d_{1}d_{2}d_{3}(d_{1}d_{2}d_{3}+2)-(d_{1}^{2}d_{2}^{2}+d_{1}^{2}d_{3}^{2}+d_{2}^{2}d_{3}^{2})]}{d_{1}^{3}d_{2}^{3}d_{3}^{3}}.
\end{split}
\end{equation}
\qed

If a tripartite state $\rho\in H_1^{d_1}\otimes H_2^{d_2}\otimes H_3^{d_3}$ is separable under the bipartition $f|gh$ ($f\neq g\neq h\in\{1, 2, 3\}$), $\rho$ can be written as
\begin{eqnarray}\label{12}
\rho=\sum_{l}p_{l}\rho_{l}^{f}\otimes \rho_{l}^{gh}, \quad p_{l}>0, \quad \sum_{l}p_{l}=1,
\end{eqnarray}
where
\begin{eqnarray}\label{13}
\rho_{l}^{f}=\frac{1}{d_{f}}I_{d_{f}}+\frac{1}{2}\sum_{(l_{f},m_{f})\neq (0,0)}t_{l_{f}m_{f}}^{f}\mathcal{Q}^{f}(l_{f},m_{f}),
\end{eqnarray}
\begin{equation}
\begin{split}\label{14}
\rho_{l}^{gh}=&\frac{1}{d_{g}d_{h}}I_{d_{g}}\otimes I_{d_{h}}+\frac{1}{2d_{h}}\sum_{(l_{h},m_{h})\neq (0,0)}t_{l_{g}m_{g}}^{g}\mathcal{Q}^{g}(l_{g},m_{g})\otimes I_{d_{h}}\\
&+\frac{1}{2d_{g}}\sum_{(l_{h},m_{h})\neq (0,0)}t_{l_{h}m_{h}}^{h}I_{d_{g}}\otimes \mathcal{Q}^{h}(l_{h},m_{h})\\
&+\frac{1}{4}\sum_{(l_{g},m_{g}),(l_{h},m_{h})\neq (0,0)}t_{l_{g}m_{g}l_{h}m_{h}}^{gh}\mathcal{Q}^{g}(l_{g},m_{g})\otimes \mathcal{Q}^{h}(l_{h},m_{h}).
\end{split}
\end{equation}
If $\rho$ is separable under the tripartition $f|g|h$, then $\rho$ is of the form
\begin{eqnarray}\label{15}
\rho=\sum_{l}p_{l}\rho_{l}^{f}\otimes \rho_{l}^{g}\otimes \rho_{l}^{h},\quad p_{l}>0,\quad \sum_{l}p_{l}=1,
\end{eqnarray}
where $\rho_{l}^{f}$ is of the form (\ref{13}) and
\begin{eqnarray}\label{16}
\rho_{l}^{g}=\frac{1}{d_{g}}I_{d_{g}}+\frac{1}{2}\sum_{(l_{g},m_{g})\neq (0,0)}t_{l_{g}m_{g}}^{g}\mathcal{Q}^{g}(l_{g},m_{g}),
\end{eqnarray}
\begin{eqnarray}\label{17}
\rho_{l}^{h}=\frac{1}{d_{h}}I_{d_{h}}+\frac{1}{2}\sum_{(l_{h},m_{h})\neq (0,0)}t_{l_{h}m_{h}}^{h}\mathcal{Q}^{h}(l_{h},m_{h}).
\end{eqnarray}

We first consider the separability of $\rho$ under bipartition $f|gh$.
Set
\begin{eqnarray}\label{18}
S(\rho_{f|gh})=\begin{pmatrix}
2& (T^{(g)})^{t}& (T^{(gh)})^{t}\\
2T^{(f)}& T^{(fg)}& T^{(fgh)}
\end{pmatrix}.
\end{eqnarray}
We have the following result. %We get the following separable theorem.

\begin{theorem}\label{thm:1}
If a tripartite quantum state $\rho\in H_1^{d_1}\otimes H_2^{d_2}\otimes H_3^{d_3}$ is separable under the bipartition $f|gh$, then
\begin{eqnarray}\label{19}
\|S_(\rho_{f|gh})\|_{tr}\leq \frac{2}{d_{f}d_{g}d_{h}}\sqrt{\frac{(d_{f}^{2}+4d_{f}-4)(d_{g}^{3}d_{h}^{3}+d_{g}^{2}d_{h}^{3}-d_{g}d_{h}^{3}+4d_{g}^{2}d_{h}^{2}-4d_{g}^{2})}{d_{g}d_{h}}}.
\end{eqnarray}
\end{theorem}
{\it Proof}~~
By (\ref{12}), (\ref{13}) and (\ref{14}), we have
\begin{equation}
\begin{split}\label{20}
&T^{(f)}=\sum_{l}p_{l}v_{l}^{f},\quad
T^{(g)}=\sum_{l}p_{l}v_{l}^{g},\quad
T^{(gh)}=\sum_{l}p_{l}v_{l}^{gh},\\
&T^{(fg)}=\sum_{l}p_{l}v_{l}^{f}(v_{l}^{g})^{t},\quad
T^{(fgh)}=\sum_{l}p_{l}v_{l}^{f}(v_{l}^{gh})^{t}.
\end{split}
\end{equation}
Applying the relations $\|A+B\|_{tr}\leq\|A\|_{tr}+\|B\|_{tr}$ and $\||a\rangle\langle b|\|_{tr}=\||a\rangle\|\||b\rangle\|$, we get
\begin{equation}
\begin{split}
\|S(\rho_{f|gh})\|_{tr}
&=\| \sum_lp_l
\left( \begin{array}{ccccccccccccccc}
            2& (v_l^g)^t& (v_l^{gh})^t \\
          2v_l^f & v_l^f(v_l^g)^t&v_l^f(v_l^{gh})^t\\
           \end{array}
      \right )
\label{A}
\|_{tr}\nonumber\\
&=\| \sum_lp_l
\left( \begin{array}{ccccccccccccccc}
              1\\
              v_l^f\\
            \end{array}
     \right )
\left( \begin{array}{ccccccccccccccc}
              2&(v_l^g)^t&(v_l^{gh})^t\\
            \end{array}
     \right )
\|_{tr}\nonumber\\
&\leq \sum_lp_l\|
\left( \begin{array}{ccccccccccccccc}
              1\\
              v_l^f\\
            \end{array}
     \right )
\left( \begin{array}{ccccccccccccccc}
              2&(v_l^g)^t&(v_l^{gh})^t\\
            \end{array}
     \right )\|_{tr}\\
&=\sum_lp_l\|
\left( \begin{array}{ccccccccccccccc}
                 1\\
               v_l^f\\
            \end{array}
     \right )\|
\|\left( \begin{array}{ccccccccccccccc}
                 2&(v_l^g)^t&(v_l^{gh})^t\\
            \end{array}
     \right )\|\\
&=\sum_lp_l\sqrt{1+\|v_l^f\|^2}\sqrt{4+\|v_l^g\|^2+\|v_l^{gh}\|^2}\\
&\leq \frac{2}{d_{f}d_{g}d_{h}}\sqrt{\frac{(d_{f}^{2}+4d_{f}-4)(d_{g}^{3}d_{h}^{3}
+d_{g}^{2}d_{h}^{3}-d_{g}d_{h}^{3}+4d_{g}^{2}d_{h}^{2}-4d_{g}^{2})}{d_{g}d_{h}}},
\end{split}
\end{equation}
where the second inequality is due to the Lemma 1 in \cite{17CCZ} and the above Lemma \ref{lemma:1}.
\qed

\begin{theorem}\label{thm:2}
If a tripartite quantum state $\rho\in H_1^{d_1}\otimes H_2^{d_2}\otimes H_3^{d_3}$ is (fully) separable under the tripartition $f|g|h$, then
\begin{eqnarray}
\|S_(\rho_{f|g|h})\|_{tr}\leq \frac{2}{d_{f}d_{g}d_{h}}\sqrt{(d_{f}^{2}+4d_{f}-4)(d_{g}^{2}d_{h}^{2}+d_{g}+d_{h}-2)},
\end{eqnarray}
where
\begin{eqnarray}\label{22}
S(\rho_{f|g|h})&=&
\left( \begin{array}{ccccccccccccccc}
            2& (T^{(g)})^{t}& (T^{(h)})^{t} \\
          2T^{(f)}& T^{(fg)}&T^{(fh)}\\
           \end{array}
      \right )
\label{A}.
\end{eqnarray}
\end{theorem}
{\it Proof}~~
By (\ref{13}), (\ref{15}), (\ref{16}) and (\ref{17}), we have
\begin{equation}
\begin{split}\label{23}
&T^{(f)}=\sum_{l}p_{l}v_{l}^{f},\quad
T^{(g)}=\sum_{l}p_{l}v_{l}^{g},\quad
T^{(h)}=\sum_{l}p_{l}v_{l}^{h},\\
&T^{(fg)}=\sum_{l}p_{l}v_{l}^{f}(v_{l}^{g})^{t},\quad
T^{(fh)}=\sum_{l}p_{l}v_{l}^{f}(v_{l}^{h})^{t}.
\end{split}
\end{equation}
Then $S(\rho_{f|g|h})$ can be expressed as follows:
\begin{eqnarray}\label{24}
S(\rho_{f|g|h})&=&\sum_lp_l
\left( \begin{array}{ccccccccccccccc}
            2& (v_l^g)^t& (v_l^{h})^t \\
          2v_l^f & v_l^f(v_l^g)^t&v_l^f(v_l^{h})^t\\
           \end{array}
      \right )
\label{A}
\nonumber\\
&=&\sum_lp_l
\left( \begin{array}{ccccccccccccccc}
              1\\
              v_l^f\\
            \end{array}
     \right )
\left( \begin{array}{ccccccccccccccc}
              2&(v_l^g)^t&(v_l^{h})^t\\
            \end{array}
     \right ),
\end{eqnarray}
from which we obtain
\begin{equation}
\begin{split}\label{25}
\|S(\rho_{f|g|h})\|_{tr}
&\leq \sum_lp_l\|
\left( \begin{array}{ccccccccccccccc}
              1\\
              v_l^f\\
            \end{array}
     \right )
\left( \begin{array}{ccccccccccccccc}
              2&(v_l^g)^t&(v_l^{h})^t\\
            \end{array}
     \right )\|_{tr}\\
&=\sum_lp_l\|
\left( \begin{array}{ccccccccccccccc}
                 1\\
               v_l^f\\
            \end{array}
     \right )\|\cdot
\|\left( \begin{array}{ccccccccccccccc}
                 2&(v_l^g)^t&(v_l^{h})^t\\
            \end{array}
     \right )\|\\
&=\sum_lp_l\sqrt{1+\|v_l^f\|^2}\sqrt{4+\|v_l^g\|^2+\|v_l^{h}\|^2}\\
&= \frac{2}{d_{f}d_{g}d_{h}}\sqrt{(d_{f}^{2}+4d_{f}-4)}\sqrt{d_{g}^{2}d_{h}^{2}+d_{g}^{2}(d_{h}-1)+d_{h}^{2}(d_{g}-1)},
\end{split}
\end{equation}
where we have used formulae $\|A+B\|_{tr}\leq\|A\|_{tr}+\|B\|_{tr}$ and $\||a\rangle\langle b|\|_{tr}=\||a\rangle\|\cdot \||b\rangle\|$.
\qed

Next we consider the genuine tripartite entanglement. For any biseparable mixture state can be written as $\rho=\sum\limits_{i}q_{i}\rho_{1}^{i}\otimes\rho_{23}^{i}+r_{i}\rho_{2}^{i}\otimes\rho_{13}^{i}+s_{i}\rho_{3}^{i}\otimes\rho_{12}^{i}$, $q_{i}\geq0, r_{i}\geq0, s_{i}\geq0$. A state is said to be genuine multipartite entangled if it can not be expressed as the convex combination of biseparable
states. Let $M_{1}(\rho)=\frac{1}{3}[\|S(\rho_{1|23})\|_{tr}+\|S(\rho_{2|13})\|_{tr}
+\|S(\rho_{3|12})\|_{tr}]$ and
\begin{equation}
\begin{split}\label{26}
M_1=&\textrm{Max}\{\frac{2}{d_{1}d_{2}d_{3}}\sqrt{\frac{(d_{1}^{2}
+4d_{1}-4)(d_{2}^{3}d_{3}^{3}+d_{2}^{2}d_{3}^{3}-d_{2}d_{3}^{3}
+4d_{2}^{2}d_{3}^{2}-4d_{2}^{2})}{d_{2}d_{3}}},\\
&\frac{2}{d_{2}d_{1}d_{3}}\sqrt{\frac{(d_{2}^{2}+4d_{2}-4)(d_{1}^{3}d_{3}^{3}
+d_{1}^{2}d_{3}^{3}-d_{1}d_{3}^{3}+4d_{1}^{2}d_{3}^{2}-4d_{1}^{2})}{d_{1}d_{3}}},\\
&\frac{2}{d_{3}d_{1}d_{2}}\sqrt{\frac{(d_{3}^{2}+4d_{3}-4)(d_{1}^{3}d_{2}^{3}
+d_{1}^{2}d_{2}^{3}-d_{1}d_{2}^{3}+4d_{1}^{2}d_{2}^{2}-4d_{1}^{2})}{d_{1}d_{2}}}\}.
\end{split}
\end{equation}
We have the following theorem.

\begin{theorem}\label{thm:3}
A mixed state $\rho\in H_{1}^{d_1}\otimes H_{2}^{d_2}\otimes H_{3}^{d_3}$ is genuine tripartite entangled if $M_{1}(\rho)>M_1$.
\end{theorem}
{\it Proof}~~
\begin{equation*}
\begin{split}
M_{1}(\rho)=&\frac{1}{3}[\|S(\rho_{1|23})\|_{tr}+\|S(\rho_{2|13})\|_{tr}
+\|S(\rho_{3|12})\|_{tr}]\\
\leq&\frac{1}{3}(\sum q_{i}\|S(\rho^{i})_{1|23}\|_{tr}+\sum r_{i}\|S(\rho^{i})_{2|13}\|_{tr}+
\sum s_{i}\|S(\rho^{i})_{3|12}\|_{tr})\\
\leq& \frac{1}{3}(M_1+M_1+M_{1})\\
=& M_1.
\end{split}
\end{equation*}
Consequently, if $M_{1}(\rho)>M_1$, $\rho$ is genuine tripartite entangled. \qed

We consider a special quantum state, if a density matrix is permutational invariant, i.e. $\rho^{p}=p\rho p^{\dagger}$, where the $p$ denotes the any permutation of the qudits. Then any biseparable permutational invariant state can be written as $\rho=\sum\limits_{i}q_{i}\rho_{1}^{i}\otimes\rho_{23}^{i}+r_{i}\rho_{2}^{i}\otimes\rho_{13}^{i}+s_{i}\rho_{3}^{i}\otimes\rho_{12}^{i}$, where $q_{i}, r_{i}, s_{i}$ are all non-zero. We can get the following corollary:
\begin{corollary}
If a density matrix is permutational invariant, then
\begin{equation*}
\begin{split}
M_{1}(\rho)=&\frac{1}{3}[\|S(\rho_{1|23})\|_{tr}+\|S(\rho_{2|13})\|_{tr}
+\|S(\rho_{3|12})\|_{tr}]\leq J_1.
\end{split}
\end{equation*}
Thus, if $M_{1}(\rho)>J_1$, $\rho$ is a genuinely entangled tripartite state, where
\begin{equation*}
\begin{split}
J_1=&\frac{1}{3}[\frac{2}{d_{1}d_{2}d_{3}}\sqrt{\frac{(d_{1}^{2}
+4d_{1}-4)(d_{2}^{3}d_{3}^{3}+d_{2}^{2}d_{3}^{3}-d_{2}d_{3}^{3}
+4d_{2}^{2}d_{3}^{2}-4d_{2}^{2})}{d_{2}d_{3}}}\\
&+\frac{2}{d_{2}d_{1}d_{3}}\sqrt{\frac{(d_{2}^{2}+4d_{2}-4)(d_{1}^{3}d_{3}^{3}
+d_{1}^{2}d_{3}^{3}-d_{1}d_{3}^{3}+4d_{1}^{2}d_{3}^{2}-4d_{1}^{2})}{d_{1}d_{3}}}\\
&+\frac{2}{d_{3}d_{1}d_{2}}\sqrt{\frac{(d_{3}^{2}+4d_{3}-4)(d_{1}^{3}d_{2}^{3}
+d_{1}^{2}d_{2}^{3}-d_{1}d_{2}^{3}+4d_{1}^{2}d_{2}^{2}-4d_{1}^{2})}{d_{1}d_{2}}}].
\end{split}
\end{equation*}
\end{corollary}

\textit{\textbf{Example 1}}
Consider $2\times 2\times 2$ quantum state
\begin{eqnarray}
\rho=\frac{x}{8}I_{8}+(1-x)|{\rm GHZ}\rangle\langle {\rm GHZ}|,\quad 0\leq x \leq 1,
\end{eqnarray}
where $|{\rm GHZ}\rangle=\frac{1}{\sqrt{2}}(|000\rangle+|111\rangle)$ and $I_{8}$ is the $8\times8$ identity matrix. By Theorem \ref{thm:1}, when $d=2$ we have
$f_{1}(x)=\|S(\rho_{f|gh})\|_{tr}-4=(2\sqrt{2}+1)(1-x)+\sqrt{(x-1)^2+4}-4$.
When $f_{1}(x)>0$, $\rho$ is not separable under the bipartition $f|gh$ for $0\leq x<0.4941$. According to the Theorem 1 in \cite{18ZZJ}, when $f_{2}(x)=-3x+4-2\sqrt{3}>0$, $\rho$ is not separable under bipartition $f|gh$ for $0\leq x<0.179$. This shows that our theorem detects more entanglement in such partition, see Fig. 1.
\begin{figure}
\centering
\includegraphics[width=8cm]{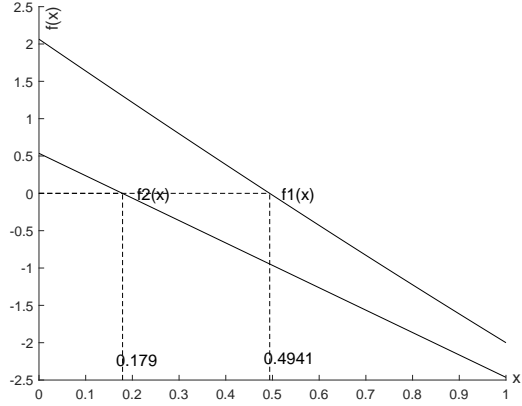}\\
\caption{The function $f_1(x)$ from Theorem \ref{thm:1} (solid curve line) and $f_2(x)$ from \cite{18ZZJ} (solid line).}
\end{figure}

From Corollary 1, we have that if $f_{3}(x)=(2\sqrt{2}+1)(1-x)+\sqrt{(x-1)^{2}+4}-4 >0$, then $\rho$ is genuine tripartite entangled for $0\leq x<0.4941$. From the lower bound of concurrence the authors in \cite{19WLLFL} show that $\rho$ is genuine tripartite entangled for $0\leq x<0.08349$. Our theorem can detect more genuine multipartite entanglement of the state.

\section{Detection of Genuine Four-partite Entanglement}

Next we consider genuine entanglement of four-partite quantum states. A four-partite state $\rho\in H_1^{d_1}\otimes H_2^{d_2}\otimes H_3^{d_3}\otimes H_4^{d_4}$ can be written as
\begin{equation}
\begin{split}
\rho=&\frac{1}{d_{1}d_{2}d_{3}d_{4}}I\otimes I\otimes I\otimes I+\sum_{f=1}^{4}\frac{d_{f}}{2d_{1}d_{2}d_{3}d_{4}}\sum t_{l_{f}m_{f}}^{f}\mathcal{Q}^{f}(l_{f},m_{f})\\
&+\sum_{1\leq f<g\leq 4}\frac{d_{f}d_{g}}{4d_{1}d_{2}d_{3}d_{4}}\sum t_{l_{f}m_{f}l_{g}m_{g}}^{fg} \mathcal{Q}^{f}(l_{f},m_{f}) \mathcal{Q}^{g}(l_{g},m_{g})\\
&+\sum_{1\leq f<g<h\leq 4}\frac{d_{f}d_{g}d_{h}}{8d_{1}d_{2}d_{3}d_{4}}\sum t_{l_{f}m_{f}l_{g}m_{g}l_{h}m_{h}}^{fgh}\mathcal{Q}^{f}(l_{f},m_{f}) \mathcal{Q}^{g}(l_{g},m_{g})\mathcal{Q}^{h}(l_{h},m_{h})\\
&+\frac{1}{16}\sum t_{l_{1}m_{1}l_{2}m_{2}l_{3}m_{3}l_{4}m_{4}}^{1234}\mathcal{Q}^{1}(l_{1},m_{1}) \mathcal{Q}^{2}(l_{2},m_{2}) \mathcal{Q}^{3}(l_{3},m_{3})\mathcal{Q}^{4}(l_{4},m_{4}),
\end{split}
\end{equation}
where ${(l_{a},m_{a})\neq (0,0)}$, $a\in \{1,2,3,4\}$, $\mathcal{Q}^{f}(l_{f},m_{f})$, $\mathcal{Q}^{g}(l_{g},m_{g})$ and $\mathcal{Q}^{h}(l_{h},m_{h})$ are operators acting on the spaces $H_{d_{f}}$, $H_{d_{g}}$ and $H_{d_{h}}$, respectively.
The coefficients are given by
$t_{l_{f}m_{f}}^{f}=\frac{2}{d_{f}}\mbox Tr\{\rho \mathcal{Q}^{f}(l_{f},m_{f})\}$, $t_{l_{f}m_{f}l_{g}m_{g}}^{fg}=\frac{4}{d_{f}d_{g}}\mbox Tr\{\rho \mathcal{Q}^{f}(l_{f},m_{f})\mathcal{Q}^{g}(l_{g},m_{g})\}$, $t_{l_{f}m_{f}l_{g}m_{g}l_{h}m_{h}}^{fgh}=\frac{8}{d_{f}d_{g}d_{h}}\mbox Tr\{\rho \\ \mathcal{Q}^{f}(l_{f},m_{f})\mathcal{Q}^{g}(l_{g},m_{g}) \mathcal{Q}^{h}(l_{h},m_{h})\}$,
$t_{l_{1}m_{1}l_{2}m_{2}l_{3}m_{3}l_{4}m_{4}}^{1234}=\frac{16}{d_{1}d_{2}d_{3}d_{4}}\mbox Tr\{\rho \mathcal{Q}^{1}(l_{1},m_{1}) \mathcal{Q}^{2}(l_{2},m_{2}) \mathcal{Q}^{3}(l_{3},\\
m_{3}) \mathcal{Q}^{4}(l_{4},m_{4})\}$. Denote
$T^{(f)}$, $T^{(fg)}$, $T^{(fgh)}$ and $T^{(1234)}$ the column vectors given by $t_{l_{f}m_{f}}^{f}$, $t_{l_{f}m_{f}l_{g}m_{g}}^{fg}$, $t_{l_{f}m_{f}l_{g}m_{g}l_{h}m_{h}}^{fgh}$ and $t_{l_{1}m_{1}l_{2}m_{2}l_{3}m_{3}l_{4}m_{4}}^{1234}$, respectively. We have
\begin{align*}
\|T^{(f)}\|^{2}&=\sum\limits(t_{l_{f}m_{f}}^{f})^{2},\\
\|T^{(fg)}\|^{2}&=\sum\limits (t_{l_{f}m_{f}l_{g}m_{g}}^{fg})^{2},\\
\|T^{(fgh)}\|^{2}&=\sum\limits(t_{l_{f}m_{f}l_{g}m_{g}l_{h}m_{h}}^{fgh})^{2},\\ \|T^{(1234)}\|^{2}&=\sum\limits(t_{l_{1}m_{1}l_{2}m_{2}l_{3}m_{3}l_{4}m_{4}}^{1234})^{2},
\end{align*}
where ${(l_{a},m_{a})\neq (0,0)}$ and $a\in \{1,2,3,4\}$.

For the four-partite quantum state $\rho \in H_{1}^{d_{1}}\otimes H_{2}^{d_{2}} \otimes H_{3}^{d_{3}}\otimes H_{4}^{d_{4}}$, we denote the bipartitions and tripartitions
as follows: $f|ghe$, $fg|he$, and $f|g|he$, where $f, g, h$ and $e$ are not equal to each other, and take from the  set $\{1, 2, 3, 4\}$.
If $\rho$ is separable under the bipartition $f|ghe$, we have
\begin{eqnarray}\label{30}
\rho_{f|ghe}=\sum_lp_l\rho_l^{(f)}\otimes\rho_l^{(ghe)},~~ p_l>0, ~~\sum_lp_l=1,
\end{eqnarray}
where
\begin{equation}
\begin{split}\label{31}
\rho_l^{f}=\frac{1}{d_f}I_{d_f}+\frac{1}{2}\sum_{(l_{f},m_{f})\neq (0,0)}t_{l_{f}m_{f}}^{f}\mathcal{Q}^{f}(l_{f},m_{f}),
\end{split}
\end{equation}
\begin{equation}
\begin{split}\label{32}
\rho^{ghe}=&\frac{1}{d_{g}d_{h}d_{e}}I_{d_{g}}\otimes I_{d_{h}}\otimes I_{d_{e}}+\sum_{g=1}^{3}\frac{d_{g}}{2d_{g}d_{h}d_{e}}\sum_{(l_{g},m_{g})\neq (0,0)}t_{l_{g}m_{g}}^{g}\mathcal{Q}^{g}(l_{g},m_{g})\otimes I_{d_{h}}\otimes I_{d_{e}}\\
&+\sum_{1\leq g<h\leq 3}\frac{d_{g}d_{h}}{4d_{g}d_{h}d_{e}}\sum_{(l_{g},m_{g}),(l_{h},m_{h})\neq (0,0)}t_{l_{g}m_{g}l_{h}m_{h}}^{gh} \mathcal{Q}^{g}(l_{g},m_{g})\otimes \mathcal{Q}^{h}(l_{h},m_{h})\otimes I_{d_{e}}\\
&+\frac{1}{8}\sum_{(l_{g},m_{g}),(l_{h},m_{h}),(l_{e},m_{e})\neq (0,0)}t_{l_{g}m_{g}l_{h}m_{h}l_{e}m_{e}}^{ghe}\mathcal{Q}^{g}(l_{g},m_{g})\otimes \mathcal{Q}^{h}(l_{h},m_{h})\otimes \mathcal{Q}^{e}(l_{e},m_{e}).
\end{split}
\end{equation}

If $\rho$ is separable under the bipartition $fg|he$, we have
\begin{eqnarray}\label{33}
\rho_{fg|he}=\sum_lp_l\rho_l^{(fg)}\otimes\rho_l^{(he)},~~ p_l>0,~~ \sum_lp_l=1,
\end{eqnarray}
where
\begin{equation}
\begin{split}\label{34}
\rho_l^{fg}=&\frac{1}{d_fd_g}I_{d_f}\otimes I_{d_g}+\frac{1}{2d_g}\sum_{(l_{f},m_{f})\neq (0,0)} t_{l_{f}m_{f}}^{f}\mathcal{Q}^{f}(l_{f},m_{f})\otimes I_{d_g}\\
&+
\frac{1}{2d_{f}}\sum_{(l_{g},m_{g})\neq (0,0)}t_{l_{g}m_{g}}^{g}I_{d_{f}}\otimes \mathcal{Q}^{g}(l_{g},m_{g})\\
&+\frac{1}{4}\sum_{(l_{f},m_{f}),(l_{g},m_{g})\neq (0,0)}t_{l_{f}m_{f}l_{g}m_{g}}^{fg}\mathcal{Q}^{f}(l_{f},m_{f})\otimes \mathcal{Q}^{g}(l_{g},m_{g}),
\end{split}
\end{equation}
\begin{equation}
\begin{split}\label{35}
\rho_l^{he}=&\frac{1}{d_hd_e}I_{d_h}\otimes I_{d_e}+\frac{1}{2d_e}\sum_{(l_{h},m_{h})\neq (0,0)} t_{l_{h}m_{h}}^{h}Q^{h}(l_{h},m_{h})\otimes I_{d_e}\\
&+
\frac{1}{2d_{h}}\sum_{(l_{e},m_{e})\neq (0,0)}t_{l_{e}m_{e}}^{e}I_{d_{h}}\otimes \mathcal{Q}^{e}(l_{e},m_{e})\\
&+\frac{1}{4}\sum_{(l_{h},m_{h}),(l_{e},m_{e})\neq (0,0)}t_{l_{h}m_{h}l_{e}m_{e}}^{he}\mathcal{Q}^{h}(l_{h},m_{h})\otimes \mathcal{Q}^{e}(l_{e},m_{e}).
\end{split}
\end{equation}
If $\rho$ is separable under the tripartitions $f|g|he$, we have
\begin{eqnarray}\label{36}
\rho_{f|g|he}=\sum_lp_l\rho_l^{(f)}\otimes\rho_l^{(g)}\otimes\rho_l^{(he)},~~ p_l>0, ~~ \sum_lp_l=1,
\end{eqnarray}
where $\rho_l^{(f)}$ and $\rho_l^{(he)}$ are given in (\ref{31}) and (\ref{35}), respectively,
\begin{equation}
\begin{split}\label{37}
\quad \rho_l^{g}=\frac{1}{d_g}I_{d_g}+\frac{1}{2}\sum_{(l_{g},m_{g})\neq (0,0)}t_{l_{g}m_{g}}^{g}\mathcal{Q}^{g}(l_{g},m_{g}).
\end{split}
\end{equation}

\begin{theorem}\label{thm:4}
If $\rho\in H_1^{d_1}\otimes H_2^{d_2}\otimes H_3^{d_3}\otimes H_4^{d_4}$ is separable under the bipartition $f|ghe$ such that $d_{i}d_{j}\geq d_{k},ijk \in \{123,132,231\}$, then
\begin{equation}
\begin{split}\label{39}
\|S_(\rho_{f|ghe})\|_{tr}\leq &\sqrt{\frac{4d_{h}^{2}+4d_{h}-4}{d_{h}^{2}}+\frac{64[d_{g}d_{h}d_{e}(d_{g}d_{h}d_{e}+2)
-(d_{g}^{2}d_{h}^{2}+d_{g}^{2}d_{e}^{2}+d_{h}^{2}d_{e}^{2})]}{d_{g}^{3}d_{h}^{3}d_{e}^{3}}}\\
&\cdot \sqrt{\frac{d_{f}^{2}+4d_{f}-4}{d_{f}^{2}}},
\end{split}
\end{equation}
where
\begin{eqnarray}\label{38}
S(\rho_{f|ghe})=
\left( \begin{array}{ccccccccccccccc}
            2&(T^{(h)})^t&(T^{(ghe)})^t\\
            2T^{(f)}&T^{(fh)}&T^{(fghe)}\\
           \end{array}
      \right )
\label{A}.
\end{eqnarray}
\end{theorem}

{\it Proof}~~
If a four-partite mixed state $\rho$ is separable under bipartition $f|ghe$, by (\ref{30}), (\ref{31}) and (\ref{32}) we have
\begin{eqnarray}\label{40}
T^{(f)}=\sum_lp_lv_l^f, \quad T^{(h)}=\sum_lp_lv_l^h, \quad T^{(ghe)}=\sum_lp_lv_l^{ghe}, \quad T^{(fghe)}=\sum_lp_lv_l^f(v_l^{ghe})^t.
\end{eqnarray}
Using Lemma 1 in \cite{17CCZ} and our Lemma \ref{lemma:2}, we have
\begin{equation}
\begin{split}
\|S(\rho_{f|ghe})\|_{tr}
=&\| \sum_lp_l
\left( \begin{array}{ccccccccccccccc}
            2&(v_l^h)^t&(v_l^{ghe})^t\\
            2v_l^f&v_l^f(v_l^h)^t&v_l^f(v_l^{ghe})^t\\
           \end{array}
      \right )
\label{A}\|_{tr}\\
=&\|\sum_lp_l
\left( \begin{array}{ccccccccccccccc}
             1\\
             v_l^f\\
            \end{array}
     \right )
\left( \begin{array}{ccccccccccccccc}
              2&(v_l^h)^t&(v_l^{ghe})^t\\
            \end{array}
     \right )\|_{tr}\\
\leq&\sum_lp_l\|
\left( \begin{array}{ccccccccccccccc}
             1\\
             v_l^f\\
            \end{array}
     \right )
\left( \begin{array}{ccccccccccccccc}
              2&(v_l^h)^t&(v_l^{ghe})^t\\
            \end{array}
     \right )\|_{tr}\\
=&\sum_lp_l\|
\left( \begin{array}{ccccccccccccccc}
             1\\
             v_l^f\\
            \end{array}
     \right )\|\cdot
\|\left( \begin{array}{ccccccccccccccc}
              2&(v_l^h)^t&(v_l^{ghe})^t\\
            \end{array}
     \right )\|\\
=&\sum_lp_l\sqrt{1+\|v_l^f\|^2}\sqrt{4+\|v_l^h\|^2+\|v_{l}^{ghe}\|^{2}}\\
\leq &\sqrt{\frac{4d_{h}^{2}+4d_{h}-4}{d_{h}^{2}}+\frac{64[d_{g}d_{h}d_{e}(d_{g}d_{h}d_{e}+2)
-(d_{g}^{2}d_{h}^{2}+d_{g}^{2}d_{e}^{2}+d_{h}^{2}d_{e}^{2})]}{d_{g}^{3}d_{h}^{3}d_{e}^{3}}}\\
&\cdot \sqrt{\frac{d_{f}^{2}+4d_{f}-4}{d_{f}^{2}}}.
\end{split}
\end{equation}\qed

\begin{theorem}\label{thm:5}
If the quantum state $\rho\in H_1^{d_1}\otimes H_2^{d_2}\otimes H_3^{d_3}\otimes H_4^{d_4}$ is separable under partition $fg|he$, then
\begin{eqnarray}
\|S(\rho_{fg|he})\|_{tr}\leq \sqrt{1+\frac{16(d_{g}^{2}-1)}{d_{f}d_{g}^{3}}}\sqrt{4
+\frac{4(d_{h}-1)}{d_{h}^{2}}+\frac{16(d_{e}^{2}-1)}{d_{h}d_{e}^{3}}},
\end{eqnarray}
where
\begin{eqnarray}
S(\rho_{fg|he})=
\left( \begin{array}{ccccccccccccccc}
            2&(T^{(h)})^t&(T^{(he)})^{t}\\
            2T^{(fg)}&T^{(fgh)}&T^{(fghe)}\\
           \end{array}
      \right )
\label{A}.
\end{eqnarray}
\end{theorem}
{\it Proof}~~
If $\rho$ is separable under partition $fg|he$, according (\ref{33}), (\ref{34}) and (\ref{35}) we have
\begin{equation}
\begin{split}
&T^{(fg)}=\sum_lp_lv_l^{fg}, ~~T^{(h)}=\sum_lp_lv_l^{h}, ~~T^{(he)}=\sum_lp_lv_l^{he}, \\
&T^{(fgh)}=\sum_lp_lv_l^{fg}(v_l^{h})^{t},~~ T^{(fghe)}=\sum_lp_lv_l^{fg}(v_l^{he})^{t}.
\end{split}
\end{equation}
Then $S(\rho_{fg|he})$ can be written as,
\begin{eqnarray}
S(\rho_{fg|he})&=&\sum_lp_l
\left( \begin{array}{ccccccccccccccc}
          2&(v_l^{h})^t&(v_l^{he})^t\\
          2v_l^{fg}&v_l^{fg}(v_l^{h})^t&v_l^{fg}(v_l^{he})^t\\
           \end{array}
      \right )
\label{A}\nonumber\\
&=&\sum_lp_l
\left( \begin{array}{ccccccccccccccc}
             1\\
             v_l^{fg}\\
            \end{array}
     \right )
\left( \begin{array}{ccccccccccccccc}
             2&(v_l^{h})^{t}&(v_l^{he})^t\\
            \end{array}
     \right ).
\end{eqnarray}
Thus,
\begin{eqnarray}
\|S(\rho_{fg|he})\|_{tr}&=&\|\sum_lp_l
\left( \begin{array}{ccccccccccccccc}
          2&(v_l^{h})^t&(v_l^{he})^t\\
          2v_l^{fg}&v_l^{fg}(v_l^{h})^t&v_l^{fg}(v_l^{he})^t\\
           \end{array}
      \right )
\label{A}\|_{tr}\nonumber\\
&\leq&\sum_lp_l\|
\left( \begin{array}{ccccccccccccccc}
             1\\
             v_l^{fg}\\
            \end{array}
     \right )
\left( \begin{array}{ccccccccccccccc}
             2&(v_l^{h})^t&(v_l^{he})^t\\
            \end{array}
     \right )\|_{tr}\nonumber\\
&=&\sum_lp_l\|
\left( \begin{array}{ccccccccccccccc}
        1\\
        v_l^{fg}\\
            \end{array}
     \right )\|\cdot
\|\left( \begin{array}{ccccccccccccccc}
          2&(v_l^{h})^t&(v_l^{he})^t\\
            \end{array}
     \right )\|\nonumber\\
&=&\sum_lp_l\sqrt{1+\|v_l^{fg}\|^2}\sqrt{4+\|v_l^{h}\|^2+\|v_l^{he}\|^2}\nonumber\\
&\leq&\sqrt{1+\frac{16(d_{g}^{2}-1)}{d_{f}d_{g}^{3}}}\sqrt{4+\frac{4(d_{h}-1)}{d_{h}^{2}}+\frac{16(d_{e}^{2}-1)}{d_{h}d_{e}^{3}}}.
\end{eqnarray}
\qed

For the tripartition $f|g|he$, we introduce
\begin{eqnarray}
S(\rho_{f|g|he})=
\left( \begin{array}{ccccccccccccccc}
           (T^{(g)})^{t}& (T^{(gh)})^{t}&(T^{(ghe)})^{t}  \\
           T^{(fg)}&T^{(fgh)}&T^{(fghe)}\\
           \end{array}
      \right )
\label{A}.
\end{eqnarray}

\begin{theorem}\label{thm:6}
If the quantum state $\rho\in H_1^{d_1}\otimes H_2^{d_2}\otimes H_3^{d_3}\otimes H_4^{d_4}$ is separable under tripartition $f|g|he$, then
\begin{eqnarray}
\|S(\rho_{f|g|he})\|_{tr}\leq\sqrt{1+\frac{4(d_{f}-1)}{d_{f}^{2}}}\sqrt{1+\frac{4(d_{h}-1)}{d_{h}^{2}}+ \frac{16(d_{e}^{2}-1)}{d_{h}d_{e}^{3}}}\sqrt{\frac{4(d_{g}-1)}{d_{g}^{2}}}.
\end{eqnarray}
\end{theorem}
{\it Proof}~~
If a four-partite mixed state $\rho$ is separable under tripartition $f|g|he$, by (\ref{31}), (\ref{35}), (\ref{36} and (\ref{37}) we have
\begin{equation}
\begin{split}
&T^{(g)}=\sum_lp_lv_l^g,~~ T^{(gh)}=\sum_lp_lv_{l}^{g}\otimes v_{l}^{h},~ ~ T^{(ghe)}=\sum_lp_lv_l^g\otimes (v_{l}^{he})^t,~~ T^{(fg)}=\sum_lp_lv_{l}^{f}(v_{l}^{g})^{t},\\
&T^{(fgh)}=\sum_lp_lv_l^f\otimes (v_{l}^{g}\otimes v_{l}^{h})^t,~~ T^{(fghe)}=\sum_lp_lv_l^f\otimes (v_l^g\otimes v_{l}^{he})^{t}.
\end{split}
\end{equation}
Thus,
\begin{equation}
\begin{split}
\|S(\rho_{f|g|he})\|_{tr}
&=\|\sum_lp_l
\left( \begin{array}{ccccccccccccccc}
           (v_{l}^{g})^{t}&(v_{l}^{g}\otimes v_{l}^{h})^{t}&(v_{l}^{g}\otimes v_{l}^{he})^{t}\\
            v_{l}^{f}(v_{l}^{g})^{t}&v_{l}^{f}(v_{l}^{g}\otimes v_{l}^{h})^{t}&v_{l}^{f}(v_{l}^{g}\otimes v_{l}^{he})^{t}\\
           \end{array}
      \right )
\label{A}\|_{tr}\nonumber\\
&\leq\sum_lp_l\|
\left( \begin{array}{ccccccccccccccc}
             1\\
             v_l^f\\
            \end{array}
     \right )
\left( \begin{array}{ccccccccccccccc}
             1&(v_{l}^{h})^{t}&(v_{l}^{he})^{t}\\
            \end{array}
     \right )\otimes (v_{l}^{g})^{t}
\|_{tr}\\
&=\sum_lp_l\|
\left( \begin{array}{ccccccccccccccc}
             1\\
             v_l^f\\
            \end{array}
     \right )\|\cdot
\|\left( \begin{array}{ccccccccccccccc}
            1&(v_{l}^{h})^{t}
             &(v_{l}^{he})^{t}\\
            \end{array}
     \right )
     \|\cdot
\|v_l^g\|
\\
&=\sum_lp_l\sqrt{1+\|v_l^f\|^2}\sqrt{1+\|v_{l}^{h}\|^2+\|v_{l}^{he}\|^2}\sqrt{\|v_{l}^{g}\|^{2}}\\
&\leq\sqrt{1+\frac{4(d_{f}-1)}{d_{f}^{2}}}\sqrt{1+\frac{4(d_{h}-1)}{d_{h}^{2}}+ \frac{16(d_{e}^{2}-1)}{d_{h}d_{e}^{3}}}\sqrt{\frac{4(d_{g}-1)}{d_{g}^{2}}}.
\end{split}
\end{equation} \qed

To deal with the genuine entanglement of four-partite quantum states, we consider the trace norm of $S$ suming over all possible bipartitions $f|ghe$ and $fg|he$, define $M_{2}(\rho)=\frac{1}{16}[\|S(\rho_{1|234})\|_{tr}+\cdot \cdot \cdot+\|S(\rho_{12|34})\|_{tr}]$. Set
\begin{equation}
\begin{split}
M_2=&\textrm{Max}\{\sqrt{\frac{d_{f}^{2}+4d_{f}-4}{d_{f}^{2}}}\sqrt{\frac{4d_{h}^{2}+4d_{h}-4}{d_{h}^{2}}
+\frac{64[d_{g}d_{h}d_{e}(d_{g}d_{h}d_{e}+2)-(d_{g}^{2}d_{h}^{2}
+d_{g}^{2}d_{e}^{2}+d_{h}^{2}d_{e}^{2})]}{d_{g}^{3}d_{h}^{3}d_{e}^{3}}},\\
& \sqrt{1+\frac{16(d_{g}^{2}-1)}{d_{f}d_{g}^{3}}}\sqrt{4+\frac{4(d_{h}-1)}{d_{h}^{2}}
+\frac{16(d_{e}^{2}-1)}{d_{h}d_{e}^{3}}}\},
\end{split}
\end{equation}
where $f,g,h$ and $e$ are not equal to each
other, and take from the set $\{1,2,3,4\}$. Using the similar method to Theorem \ref{thm:3}, we have
\begin{theorem}\label{thm:7}
For any four-partite state $\rho\in H_{1}^{d_{1}}\otimes H_{2}^{d_{2}}\otimes H_{3}^{d_{3}}\otimes H_{4}^{d_{4}}$ such that $d_{i}d_{j}\geq d_{k},ijk \in \{123,132,231\}$, if $M_{2}(\rho)>M_2$ then $\rho$ is genuine four-partite entangled.
\end{theorem}

If the density matrix is permutational invariant, we can get the following corollary:
\begin{corollary}
If a density matrix is permutational invariant, then $M_{2}(\rho)\leq J_2$, if $M_{2}(\rho)> J_2$, $\rho$ is a genuinely entangled tripartite state, where
\begin{equation*}
\begin{split}
J_2=&\frac{1}{16}[\sum\sqrt{\frac{d_{f}^{2}+4d_{f}-4}{d_{f}^{2}}}\sqrt{\frac{4d_{h}^{2}+4d_{h}-4}{d_{h}^{2}}
+\frac{64[d_{g}d_{h}d_{e}(d_{g}d_{h}d_{e}+2)-(d_{g}^{2}d_{h}^{2}
+d_{g}^{2}d_{e}^{2}+d_{h}^{2}d_{e}^{2})]}{d_{g}^{3}d_{h}^{3}d_{e}^{3}}}\\
&+\sum \sqrt{1+\frac{16(d_{g}^{2}-1)}{d_{f}d_{g}^{3}}}\sqrt{4+\frac{4(d_{h}-1)}{d_{h}^{2}}
+\frac{16(d_{e}^{2}-1)}{d_{h}d_{e}^{3}}}],
\end{split}
\end{equation*}
the first $\sum$ sums over all possible bipartitions $f|ghe$, the second $\sum$ sums over all possible bipartitions $fg|he$.
\end{corollary}

\textit{\textbf{Example 2}}
Consider the four-qubit quantum state $\rho\in H_1^2\otimes H_2^2\otimes H_3^2\otimes H_4^2$,
\begin{eqnarray}
\rho=x|\psi\rangle\langle\psi|+\frac{1-x}{16}I_{16},
\end{eqnarray}
where $|\psi\rangle=\frac{1}{\sqrt{2}}(|0000\rangle+|1111\rangle)$ and $I_{16}$ is the $16\times 16$ identity matrix.

For bipartition $f|ghe$, we have $f_4(x)=\|S(\rho_{f|ghe})\|_{tr}-\sqrt{18}=(4+\sqrt{2})x+2-\sqrt{18}>0$ by Theorem \ref{thm:4}. We also get  $f_5(x)=\|S(\rho_{f|g|he})\|_{tr}-\sqrt{10}=(\sqrt{2}+5)x-\sqrt{10}$ under tripartition $f|g|he$ by Theorem \ref{thm:6}. By Theorem 3 of Ref. \cite{20LWSF}, $f_6(x)=9x^{2}-4$ and $f_{7}(x)=9x^{2}-3$ for $f|ghe$ and $f|g|he$ can be obtained. The entanglement $\rho$ in different partitions is shown
in Table 1. Fig. 2 shows that our method can detect more entangled states.
\begin{table}[h]
\begin{center}
\begin{minipage}{380pt}
\caption{The entanglement range of $\rho$ in different partition}\label{tab1}%
\begin{tabular}{l c c }
\toprule
 & the range of entanglement for $f|ghe$ & the range of entanglement for $f|g|he$\\
\midrule
    &$f_{4}(x)>0,$\ \  $0.4142<x\leq1$   & $f_{5}(x)>0,$\ \ $ 0.493<x\leq1$  \\
 & $f_{6}(x)>0,$\ \  $0.6667<x\leq1$    & $f_{7}(x)>0, $\ \  $0.5774<x\leq0.6667$  \\
\bottomrule
\end{tabular}
\end{minipage}
\end{center}
\end{table}
\begin{figure}
\centering
\includegraphics[width=15cm]{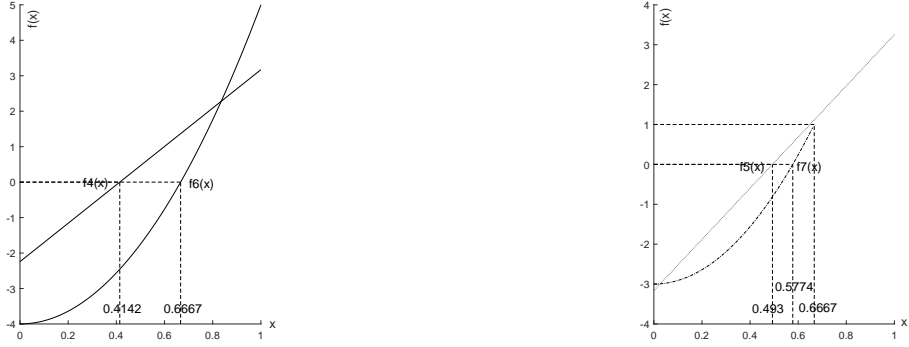}\\
\caption{ The function $f_{4}(x)$ from Theorem \ref{thm:4} (solid line), $f_{6}(x)$ from Theorem 3 in \cite{20LWSF} (solid curve line), $f_{5}(x)$ from Theorem \ref{thm:6} (dashed line) and $f_{7}(x)$ from Theorem 3 in \cite{20LWSF} (dash-dot curve line).}
\end{figure}
Moreover, from our Corollary 2, $M_{2}(\rho)=\frac{1}{4}[(4+\sqrt{2})x+2]+\frac{3}{4}[4x
+\sqrt{-\sqrt{(x^{2}+2x+2)(5x^{2}-2x+2)}+3x^{2}+2}
+\sqrt{\sqrt{(x^{2}+2x+2)(5x^{2}-2x+2)}+3x^{2}+2}]$, $\rho$ is genuine four-qubit entangled for $0.6361<x\leq 1$.
However, this genuine four-qubit entanglement can not be considered by the criterion given in \cite{20LWSF}.

{\bf Remark} The authors \cite{17CCZ} studied the separability for bipartite quantum systems by all correlation tensors and only gave a necessary condition of fully separable for multipartite quantum states. While in our current approach, we study the entanglement under different partition and present fully separable, bi-separable and tri-separable necessary conditions in tripartite and four-partite quantum systems. Moreover, we derive the genuine multipartite entanglement criteria of tripartite and four-partite quantum systems. Our main structural matrices in the algorithm differ from the Ref.\cite{17CCZ} both in structure and in form. We only use a part of correlation tensors in the HW representation to study the entanglement and genuine multipartite entanglement. Detailed examples show that our criteria can detect more entanglement than previous studies.

\section{ Conclusion}
We have studied the entanglement and genuine multipartite entanglement in tripartite and four-partite quantum systems. Taking the advantage of the HW representation, we have derived the upper bounds on the norms of correlation tensors and the separability criteria under any partitions. Detailed examples show that our criteria are able to detect genuine multipartite entanglement more effectively than some existing criteria. Our approach can be also applied to more general multipartite systems.

\noindent\textbf{Acknowledgments} This work is supported by the National Natural Science Foundation of China under
grant nos. 11101017, 11531004, 11726016, 12075159, 12126351 and 12171044, Simons Foundation under grant no. 523868,
Beijing Natural Science Foundation (grant no. Z190005), Academy for Multidisciplinary Studies, Capital
Normal University, and Shenzhen Institute for Quantum Science and Engineering, Southern University
of Science and Technology (no. SIQSE202001), and the Academician Innovation Platform of Hainan Province.

\noindent\textbf {Data Availability Statements} All data generated or analysed during this study are available from the corresponding author on reasonable request.


\begin{thebibliography}{99}
\bibitem{1EA} Ekert, A. K.: Quantum cryptogaphy based on Bell's theorem. Phys. Rev. Lett. {\bf 67}, 661 (1991)

\bibitem{2BW} Bennett, C. H., Wiesner, S. J.: Communication via one-and two-particle operators on Einstein-Podolsky-Rosen states. Phys. Rev. Lett. {\bf 69}, 2881 (1992)

\bibitem{3BBJ} Bennett, C. H., Brassard, G., Jozsa, R.: Teleporting an unknown quantum state via dual classical and Einstein-
Podolsky-Rosen channels. Phys. Rev. Lett. {\bf 70}, 1895 (1993)

\bibitem{4HHH} Peres, A.: Phys. Rev. Lett. \textbf {77}, 1413 (1996)

\bibitem{5B} Terhal, B.: Phys. Lett. A {\bf 271}, 319 (2000)

\bibitem{6CW} Chen, K., Wu, L. A.: A matrix realignment method for recognizing entanglement. Quantum Inf. Comput. {\bf 3}, 193 (2002)

\bibitem{7R} Rudolph, O.: Some properties of the computable cross-norm criterion for separability. Phys. Rev. A {\bf 67}, 032312 (2003)

\bibitem{8RO} Rudolph, O.: Further results on the cross norm criterion for separability. Quantum Inf. Process. {\bf 4}, 219 (2005)

\bibitem{9LWSCF} Li, M., Wang, J., Shen, S., Chen, Z., Fei, S. M.: Detection and measure of genuine tripartite entanglement with partial transposition and realignment of density matrices. Sci. Rep. {\bf 7}, 17274 (2018)

\bibitem{10CMCS} Chen, Z. H., Ma, Z. H., Chen, J. L., Severini, S.: Improved lower bounds on genuine-multipartite-entanglement concurrence. Phys. Rev. A {\bf 785}, 062320 (2012)

\bibitem{11ZFFW} Zhao, H., Fei, S. M., Fan, J., Wang, Z. X.: Inequalities detecting entanglement for arbitrary bipartite systems. Int. J. Quantum Inform. {\bf 12}, 1450013 (2014)

\bibitem{12HS} Huber, M., Sengupta, R.: Witnessing genuine multipartite entanglement with positive maps. Phys. Rev. Lett. {\bf 113}, 100501 (2014)

\bibitem{13SYLF} S.~Q. Shen, J. Yu, M.~Li and S.~M.~Fei, Scientific Reports, 6, 28850 (2016)

\bibitem{14V} Vicente, J.: Separability criteria based on the bloch representation of density matrices. Quantum Inf. Comput. {\bf 7}, 624 (2006)

\bibitem{15BK} Bertlmann, R. A., Krammer, P.: Bloch vectors for qudits. J. Phys. A: Math.
Theor. {\bf 41}, 235303 (2008)

\bibitem{16AEHK} Asadian, A., Erker, P., Huber, M., Kl\"{o}ckl, C.: Heisenberg-Weyl Observables: Bloch vectors in phase space. Phys. Rev. A {\bf 94}, 010301 (2016)

\bibitem{17CCZ} Chang, J., Cui, M., Zhang, T., Fei, S. M.: Separability criteria based on Heisenberg-Weyl representation of density matrices. Chinese Physics B. {\bf 27}, 030302, (2018)

\bibitem{18ZZJ} Zhao, H., Zhang, M. M., Jing, N., Wang, Z. X.: Separability criteria based on Bloch representation of density matrices. Quantum Inf. Process. {\bf 19}, 1 (2020)

\bibitem{19WLLFL} Wang, J., Li, M., Li, H., Fei, S. M., Li-Jost, X.: Bounds on multipartite concurrence and tangle. Quantum Inf. Process. {\bf 15}, 4211 (2016)

\bibitem{20LWSF} Li, M., Wang, Z., Wang, J., Shen, S., Fei, S. M.: The norms of Bloch vectors and classification of four-qudits quantum states. EPL (Europhysics Letters). {\bf 125}, 20006 (2019)
\end{thebibliography}
\end{document}